\documentclass[multphys,vecphys]{svmult}

\usepackage{makeidx}     
\usepackage{graphicx}    
\usepackage{multicol}    

\makeindex             

\begin{document}

\title*{The Iron Abundance and Density Structure of the Inner Ring around 
SN 1987A}
\titlerunning{the Inner Ring around SN 1987A}
\author{S.Mattila\inst{1}\and
P.Lundqvist\inst{1}\and
P.Meikle\inst{2}\and
R.Stathakis\inst{3}\and
R.Cannon\inst{3}}
\authorrunning{Mattila et al.}
\institute{Stockholm Observatory, AlbaNova, Dept. of Astronomy, Stockholm 
SE-106 91, Sweden \texttt{seppo@astro.su.se}
\and Blackett Laboratory, Imperial College, Prince Consort Road, London SW7 2BW, UK
\and Anglo-Australian Observatory, PO Box 296, Epping, NSW 1710, Australia}
%
%
\maketitle

\renewcommand{\abstractname}{Abstract}
\begin{abstract}
We present a spectroscopic study of the inner circumstellar ring around 
SN 1987A. The aim is to determine the elemental abundances and density 
structure, with particular emphasis on the abundance of iron. We 
acquired and analysed optical spectra at the Anglo-Australian Telescope (AAT) 
between 1400 and 4300 days post-explosion. We also assembled from the 
literature all available optical/near-IR line fluxes of the inner ring. The 
observed line light curves were then compared with a photoionisation model for 
the inner ring. This indicates an iron abundance of  (0.20$\pm$0.08) $\times$ 
solar which is lower than that generally seen in the Large Magellanic Cloud 
(LMC).
\end{abstract}

\section{Motivation}
X-ray observations [2] indicate an iron abundance of 0.1 $\times$ solar for the 
circumstellar medium (CSM) of SN 1987A. This is surprisingly low compared to the iron abundances, ranging
between $\sim$0.25 and $\sim$0.50 $\times$ solar, e.g. [7,9,1], generally observed in the LMC. 
Such a large under-abundance of iron could result from the depletion of iron on dust grains in the red 
supergiant wind of the progenitor star before the supernova (SN) exploded. 

\section{The Observed Line Light curves for the Inner Ring}

\subsection{AAT Observations of SN 1987A Inner Ring}
Optical spectroscopic observations of SN 1987A were carried out using the 
Royal Greenwich Observatory (RGO) Spectrograph on the AAT between 1991 and 
1998. Data were obtained at four different epochs: 1416, 1680, 1991, and 4309 
days post-explosion. At each epoch, the observations comprised a brief, wide 
slit integration (5-15 min; 5.3"-10.0"), and a longer duration, narrow slit 
integration (1-3 hours; 1.5"-2.0").  The SN observations spanned air masses $\sim$1.5 to 
$\sim$3.0 (i.e. zenith distances of 50 to 70 degrees), making accurate flux 
calibration quite challenging.  We therefore carried out the observations with 
the slit position angle (PA) set to be roughly the same as the line joining 
Stars 2 and 3.  This meant that Star 2 lay well within the broad slit and so 
could be used to correct for the variable transmission of the atmosphere 
(for details see [14]).  However, an additional problem was that, 
since the slit PA was generally not at the parallactic angle, when the narrow 
slit was used, 
atmospheric refraction could introduce wavelength-dependent vignetting of the 
ring spectra.  Moreover, the magnitude of this effect 
differed from that experienced by Star 2 since the latter lay nearer the edge 
of the slit. Consequently, the fluxing uncertainty introduced could be as 
large as $\pm$40$\%$.

\subsection{Star 3 Contamination Removal Using HST Archival Data}
The AAT observations suffered from seeing ranging between $\sim$1" and $\sim$3" and so 
the inner ring spectra were always significantly contaminated by light from 
the two nearby stars.  In general, the CSM emission lines could 
be easily distinguished from the continuum originating from the SN and the two 
stars.  However, Star 3 is a Be star showing strong H$\alpha$ and H$\beta$ 
emission together with optical variability of $\sim$0.5 magnitudes [22].  
Consequently, it could affect significantly the observed line fluxes from the 
inner ring, especially at the later epochs when the ring was fainter.  To 
assess and correct for possible Star 3 contamination, we searched the HST 
archive for suitable optical spectroscopic (STIS) and photometric (WFPC2) 
observations.

STIS spectra with a large enough slit aperture (52"x2") and suitable centering 
and orientation to include Star 3 were selected from the HST archive. These 
spectra confirm the findings of Wang et al. [22] that there are indeed no 
forbidden lines present in the Star 3 spectrum (see [14]).
The vignetted narrow slit spectra were therefore scaled in flux to match those 
of the unvignetted broad slit spectra using the brightest forbidden lines 
present in both the spectra. Deriving line fluxes for the whole inner ring 
this way assumes homogeneous ring geometry.  WFPC2 F502N and 
F656N images which were contemporary with our AAT spectra were then selected 
from the HST archive. Average H$\alpha$ and H$\beta$ fluxes of 44 and 13 
$\times$ 10$^{-15}$ erg s$^{-1}$ cm$^{-2}$, respectively, were found for Star 3, 
with a maximum variability of $\sim$20$\%$ over a 6 year time span.  These 
fluxes were used to correct for Star 3 contamination of the hydrogen line 
fluxes in the broad slit AAT observations. In 
addition, the accuracy of the absolute fluxing of our AAT observations was 
checked using the HST inner ring data. This indicates that our ground-based 
fluxes are consistent with the HST measurements to within the estimated 
errors of $\pm$20-40$\%$.

\subsection{The Line Light Curves}
From the AAT spectra, we measured optical line fluxes at 1416, 1680, 2864 and 
4309 days for H, He, N, O, Ne, S, Ar, Ca and Fe (e.g. Fig.1). We also assembled from the 
literature all available optical/near-IR line fluxes of the inner ring.  This 
includes data from the following sources: [19] (307 days), [20] (511, 552, 586, 678, and 735 days), 
[15] (574, 668, 695, 734, 840, 958, and 1114 days), [16] (1050 days), [21] (1280 days), 
[3] (1344 days), [5] (1348, 1469, 1734, 1822, and 2122 days).  
The resulting dataset (see [14]) spans $300-4300$ days post-explosion, 
i.e., from the time when the ring first became visible to the time when the first signs 
of ejecta/ring collision were seen at these wavelengths. 

\subsection{Modelling the Line Light Curves of the Inner Ring}
The latest version of the photoionisation code described in [10, 11, 12]
was used to model the emission line light curves of the inner 
ring. In the model the ring is initially ionised by the soft X-ray and UV 
photons emitted in the SN shock break-out. The SN flash with the temporal and 
spectral characteristics of the 500full1 model [4] sets up the initial ionisation 
structure of the ring, and the gas then recombines and cools. To model the emission line fluxes 
of iron, we collected from the literature the latest atomic data for the most 
abundant ions of iron in the ring (see [14]). Thanks to the Iron 
Project [8] these atomic data have improved significantly during 
the last few years and now make such abundance determinations more reliable.

\begin{figure}
\centering
\includegraphics[height=20cm]{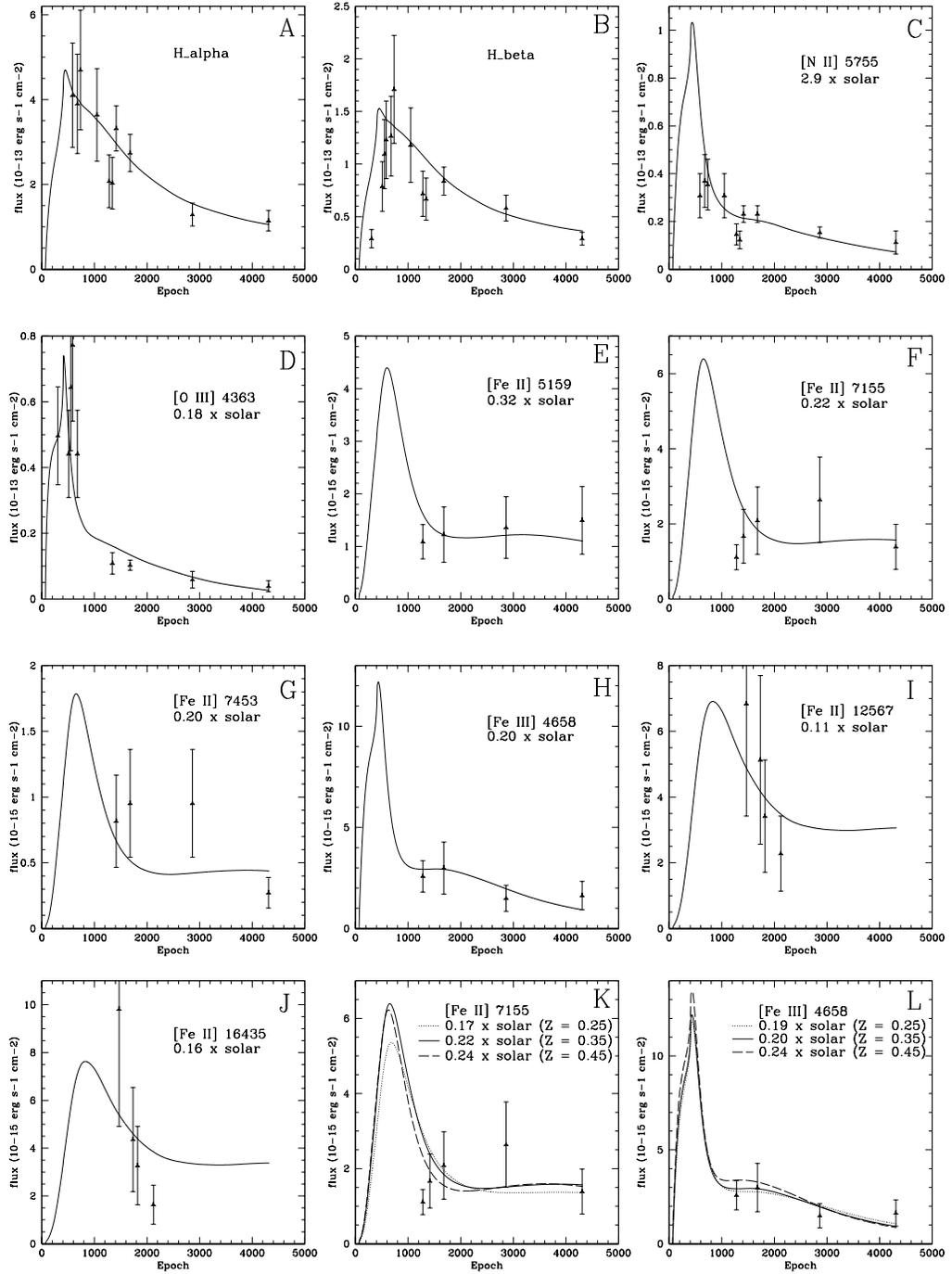}
\caption{$\bf{A-J:}$ The observed fluxes of lines of H, N, O, and Fe 
plotted together with the model fluxes. The observed line fluxes have been 
dereddened assuming
E$_{B-V}$ of 0.05 and 0.15 for the Galaxy and the LMC, respectively. The
abundance used for each model fit is given relative to
solar. $\bf{K-L:}$ The observed line fluxes of [Fe II] 7155\AA~and 
[Fe III] 4658\AA~plotted together with the model fluxes for three 
different metallicities.}
\label{fig:1}       
\end{figure}

\subsection{The Density Structure}
The mass and density structure of the ionised gas in the inner ring was 
determined by searching for satisfactory model fits for (1) the absolute line 
fluxes of H$\alpha$ and H$\beta$ (Fig.1A-B), and (2) the observed time 
evolution of the line fluxes of all the other elements (e.g., Fig.1C-D). We 
found that five density components were needed: 10$^{3}$, 2 $\times$ 10$^{3}$, 3 $\times$ 
10$^{3}$, 2 $\times$ 10$^{4}$, and 4 $\times$ 10$^{4}$ atoms 
cm$^{-3}$ with masses of ionised gas of 0.4, 1.1, 1.0, 0.8, and 0.5 $\times$ 
10$^{-2}$ M$_{\odot}$, respectively. The low-density components dominate the line light
curves at the later times. High resolution (HST) images indicate that this
low-density gas is situated, on average, further away from the SN than the gas with a
higher density [13]. In reality there is also probably a continuous range of densities 
within the ring rather than a few discrete density components. 

\subsection{The Iron Abundance}
The elemental abundances yielding the observed absolute line fluxes were then 
determined (Fig.1E-J). This was done by altering the total number of emitting 
ions relative to hydrogen. However, we have not yet included the effects of 
these modified abundances on the temperature and ionisation structure of the 
ring.  Therefore, the abundances derived for the more abundant elements, 
e.g. He, N, O, should be considered as only preliminary results (e.g. 
Fig.1C-D).  However, the effect of the modified iron abundance on the temperature 
and ionisation structure is not significant, thus allowing its robust determination.

Our estimate of the iron abundance is based on five [Fe II] and one [Fe III] 
lines (Fig.1E-J). This indicates an average abundance of 0.20 
times solar with a dispersion of 35$\%$. In addition, we estimated 
the effects of the assumed He, N, and O abundances and the overall metallicity 
on the determined iron abundance. We found that the iron model fluxes are 
rather insensitive to the He/H and N/O ratios. However, the overall 
metallicity has a much larger effect on the cooling, and thus on the iron 
model line light curves (Fig.1K-L). We estimate that this introduces an uncertainty of 
15$\%$ in the average iron abundance. Therefore, from this study, we deduce an iron 
abundance of (0.20 $\pm$ 0.08) times solar (see [14]). This abundance 
is larger by a factor of two relative to that found by Borkowski, Blondin $\&$ 
McCray [2] from X-ray observations. However, it is at the lower end of the iron abundance 
range (0.25-0.50 times solar) found for different locations within the LMC. We note that also
an inner ring silicon abundance of a factor of $\sim$2 lower 
than normal for the LMC was found recently by Lundqvist $\&$ Sonneborn [13]. These low 
abundances are probably due to depletion onto dust grains. The existence of dust in the 
ring material has been demonstrated by mid-IR observations [17, 6].

\vspace{+0.2cm}

We thank L. Smith, R. Terlevich, and R. Cumming for helpful discussions.
%
%

%
%



\printindex
\end{document}